\documentclass{ctr_summer}

\usepackage{ctrfont}

\usepackage{epsfig}
\usepackage{psfrag}
\usepackage{amsmath,rotating}

\title{Problems of astrophysical turbulent convection:\\ thermal convection in a layer without boundaries}
\shorttitle{Thermal convection without boundaries}

\author{ R.~D.~Simitev\footnote{School of Mathematics and Statistics,
    University of Glasgow} \and F.~H.~Busse\footnote{Institute of
    Physics, University of Bayreuth} 
  }

\shortauthor{ R.~D.~Simitev \& F.~H.~Busse}

\begin{document}

\setcounter{page}{485}

\maketitle

Thermal convection in fluid layers heated from below are usually
realized experimentally as well as treated theoretically with fixed
boundaries on which conditions for the temperature and the velocity
field are prescribed. The thermal and velocity boundary layers
attached to the upper and lower boundaries determine to a large extent
the properties of turbulent convection at high Rayleigh numbers. Fixed
boundaries are often absent in natural realizations of thermal
convection. This paper studies the properties of convection driven by a
planar heat source below a cooling source of equal size immersed in an
otherwise stably stratified fluid layer are studied in this
paper. Unavoidable boundaries do not influence the convection flow
since they are separated from the active convection layer by nearly
motionless stably stratified regions. The onset of convection occurs
in an inner unstably stratified region where the mean temperature
gradient is reversed. But the region of a reversed horizontally
averaged temperature gradient disappears at higher amplitudes of
convection such that the vertical derivative of the mean temperature
no longer changes its sign.
\vskip0.1in
\hrule

\section{Introduction}
\label{s:intro}

High Rayleigh number thermal convection has received much attention in
the past decades for several reasons. On the one hand, thermal
convection represents the most common fluid flow in nature and the
understanding of its properties is essential for assessing energy
transports in planetary atmospheres and in stars. On the other hand,
turbulent convection has become the paradigmatic example for the study
of turbulent transports. The experimental investigations are
facilitated by the fact that unlike the situation in channel flows, the
advection of turbulent eddies by a mean flow is absent. Moreover, the
temperature as a scalar quantity offers additional opportunities for
quantitative observations. These properties have also motivated theoretical
studies, and numerous computational simulations have appeared in the
recent literature. For an overview of many aspects of convection we
refer to the recent book by \cite{Lappa:2010}.

A common property of laboratory convection experiments as well as of
theoretical formulations of convection problems is the assumption of
fixed boundaries on which the temperature is prescribed and the
velocity field usually must vanish. As a result, turbulent convection
is determined to a large extent by the properties of the thermal and
velocity boundary layers forming at the fixed upper and lower
boundaries. This situation is not typical for most convection flows
realized in nature. Often convection is driven by heat sources owing
to absorbed radiation in conjunction with cooling by emitted
radiation. An example for such a convection system is the solar
convection zone in which heat is partly transported by convection in
the region  where a purely radiative transport would require an
unstable superadiabatic gradient of entropy. 

In this paper we are not interested in studying any realistic
convection layer. Instead, we are formulating a theoretical problem in
which convection without influence from boundaries can be investigated
in a simple setting. We have chosen the case of a heat source layer
below a cooling layer of equal size. Both are embedded in a extended,
stably stratified fluid layer such that the velocity field is strongly
damped above and below the heating and cooling layers. The unavoidable
boundaries introduced for the numerical analysis thus exert only a
minimal influence on the convection flow.     

In the following section the mathematical problem is
described. Numerical simulation of two-dimensional convection for
different Prandtl numbers is carried out and the results are 
discussed in Section 3. 
An outlook on future work is given in Section 4.

\section{Mathematical formulation of the problem}

We consider a fluid layer of height $h$ and adopt the Boussinesq
approximation, i.e., all material properties are assumed to be constant
except for the temperature dependence of the density which is taken
into account only in connection with the gravity term. Using as scales
the length $h$, the time $h^2 / \kappa$,  where $\kappa$ is the
thermal diffusivity of the fluid, and the temperature 
$q h^2/c\kappa$, where $q$ is a heat source density and $c$ is the
specific heat of the fluid, we obtain the dimensionless equations of 
motion for the velocity vector $\vec u$ and the heat equation for the
deviation  $\Theta$ from the static temperature distribution, $T_s$
\begin{subequations}
\begin{gather}
\begin{align}
(\partial_t \vec{u} + \vec u \cdot \nabla \vec u)/P 
& = - \nabla \pi + R\Theta \vec k + \nabla^2 \vec u , \\
\nabla \cdot \vec u & = 0, \\
\partial_t \Theta + \vec u \cdot \nabla \Theta& = u_z T'_s + \nabla^2 \Theta, 
\end{align}
\end{gather}
\end{subequations}
where $\partial_t$ denotes the partial derivative with respect to time
$t$ and where all terms in the equation of motion that can be written
as gradients have been combined into $ \nabla \pi$. The dimensionless
parameter, the Rayleigh number $R$ and the Prandtl number $P$ are
given by 
\begin{equation}
R = \frac{\alpha g q d^6}{c\nu \kappa}, \qquad P = \frac{\nu}{\kappa},
 \end{equation}
where $\alpha$ is the coefficient of thermal expansion, $g$ is gravity
and $\nu$ is the kinematic viscosity. We shall use a cartesian system
of coordinates with the $z$-coordinate and the unit vector $\vec k$ in
the direction opposite to gravity. 
A heat source distribution
antisymmetric with respect to $z=0$ is chosen such that the static
temperature distribution, $T_s(z)$, is governed by the equation
\begin{equation}
\frac{d^2T_s}{dz^2}= \frac{2\gamma \tanh \gamma z}{(\cosh \gamma z)^2},
\end{equation}
where $\gamma$ is assumed to be large in comparison to unity such that
heat and cooling sources are finite only close to $z=0$ and decay
rapidly towards the boundaries at $z=\pm1/2$.  
Integration of Eq.~(2.3) yields
\begin{equation}
\frac{dT_s}{dz}= \beta -\frac{1}{(\cosh \gamma z)^2} \quad \text{and} \quad T_s(z)= \beta z -\frac{1}{\gamma}\tanh \gamma z,
\end{equation}
where $\beta$ measures the stable stratification.  We shall use
stress-free boundary conditions and require that the $z$-derivative of
$\Theta$ also vanishes, 
\begin{align}
&
u_z = \partial^2_{zz}u_z = \partial_z \Theta = 0 
&
\enspace \mbox{ at }
\enspace z=\pm \frac{1}{2}.
\end{align}

In solving the problem described by Eqs.~(2.1) and (2.5) we start
with the two-dimensional case in which the velocity field can be
described by a stream function, 
\begin{equation}
u_z = \frac{\partial \phi}{\partial x}, \qquad u_x = -\frac{\partial \phi}{\partial z}.
 \end{equation}
The equations for the stream function $\phi$, the vorticity $V$ and
deviation $\Theta$ of the temperature from its static distribution can
now be written in the form 
\begin{subequations}
\begin{gather}
V=-\nabla^2 \phi, \\
(\partial_t V -\partial_z \phi \partial_x V + \partial_x \phi \partial_z V)/P - \nabla^2 V= -R\partial_x \Theta,\\
\partial_t \Theta -\partial_z \phi \partial_x \Theta + \partial_x \phi \partial_z \Theta  = -\partial_x \phi T'_s + \nabla^2 \Theta, 
\end{gather}
\end{subequations}
which can be solved more easily than the original equations since the
pressure gradient has been eliminated. In the horizontal $x$-direction
periodic boundary conditions will be applied at $x=\pm \Gamma/2$.
The boundary conditions (Eq.~2.5) now assume the form 
\begin{align}
&
\phi = V = \partial_z \Theta = 0 
&
\enspace \mbox{ at }
\enspace z=\pm \frac{1}{2}.
\end{align}
In the following analysis we shall restrict attention to the case
$\beta = 0.5$ and $\gamma = 10$ which is representative for a narrow
convection layer imbedded in a wider stably stratified layer. 

\begin{figure}
\vspace*{7mm}
\psfrag{H}{$H$}
\psfrag{R}{$R$}
\psfrag{a}{$a$}
\begin{center}
\epsfig{file=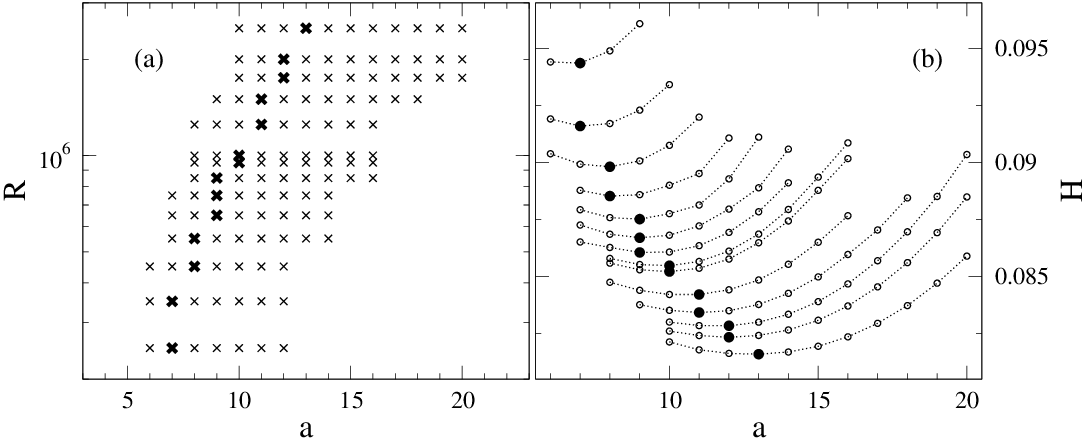,width=\textwidth,clip=}
\end{center}
\caption{Range of  wavenumbers of convection rolls for $P=1$, $\gamma=10$,
   $\beta=0.5$.
  (a)
The range of possible roll wavenumbers $a$ for a given
  Rayleigh number $R$, with the dominant one indicated by a thick
  cross.
  (b)
  The value of $H\equiv\beta/2-T(1/2)$ as a function of the values
  of possible roll wavenumbers $a$. The dominant wavenumber is
  that which minimizes $H$. The values of $R$ in (b)  are the same
   as those in (a).}
\vspace{7mm}
\psfrag{H}{$H$}
\psfrag{R}{$R$}
\psfrag{adom}{dom$(a)$}
\begin{center}
\epsfig{file=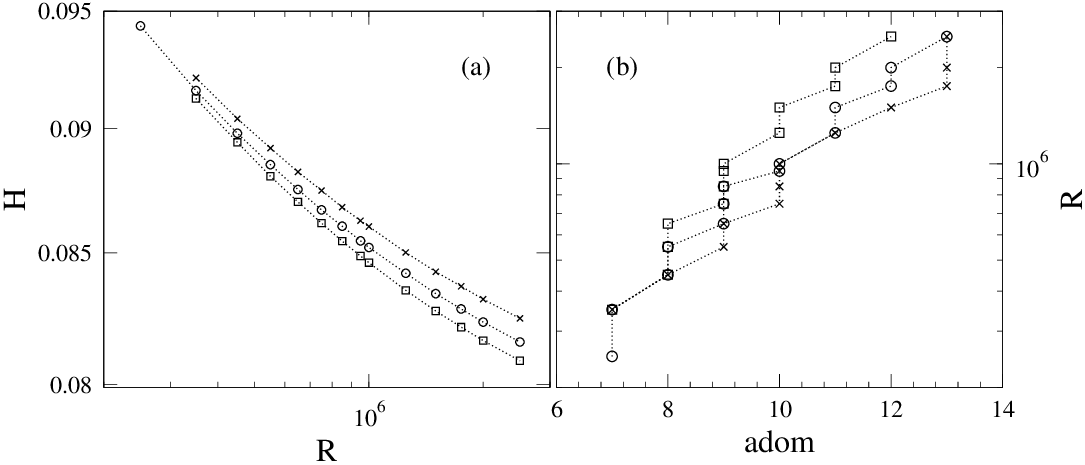,width=\textwidth,clip=}
\end{center}
\caption{(a) The quantity $H\equiv\beta/2-T(1/2)$ at the upper
  boundary, and (b) the dominant wavenumber versus the Rayleigh
  number $R$ for $P=0.5$ (crosses), $P=1$ (circles), $P=10$ (squares)
  in the case $\gamma=10$, $\beta=0.5$.
  }
\end{figure}

For the numerical solution of Eqs.~(2.7) and (2.8) we have
adopted the finite-element method as implemented in the
commercially-available software platform COMSOL v.~3.5
\cite{COMSOL:2010}
For the spatial discretization we have used regular rectangular meshes
typically consisting of 19200 elements and Lagrange shape functions.
For the time integration we have used a backward-difference formula of
order 4 with adaptive step control. The relative and the absolute error
tolerances have been set to $10^{-6}$ and $10^{-8}$, respectively.

\section{Two-dimensional convection}

The critical Rayleigh number $R_c$ for onset of convection and the
corresponding wavenumber $a_c$ in  the case $\beta = 0.5$ and $\gamma
= 10$ can  be determined numerically with a shooting method. The
result 
\begin{equation}
R_c = 144095, \qquad a_c = 5.86
\end{equation}
indicates that for $R>R_c$ convection rolls with the wavelength
$2\pi/\alpha_c$ grow and become asymptotically steady solutions as
verified by the nonlinear analysis described below. The relatively
high values (Eq.~3.1) -- as compared with the values $R_c=6.75\pi^4$ with
$\alpha_c=\pi/\sqrt{2}$ for the corresponding Rayleigh-B\'enard
problem -- reflect the reduction of the height of the convecting region
from the total height of the layer. The result is not sensitive to the
applied boundary conditions (Eq.~2.8). When the thermal boundary condition
is replaced by $\Theta = 0$ at $z=\pm \frac{1}{2}$, the result
\begin{equation}
R_c = 145154, \qquad a_c = 5.90
\end{equation}
is obtained.

As the Rayleigh number increases beyond the critical value, the
preferred wavenumber of finite amplitude convection increases as the
wavelength of convection assumes values corresponding to the typical
thickness of the order of $1/\gamma$ of the most strongly convecting
part of the layer. As indicated in Fig.~1(a), there is a finite spread
of wavenumbers that can be realized at supercritical Rayleigh
numbers. Here, all integer values have been indicated for which
computations with $\Gamma = 2\pi/a$ gave solutions with just a single
wavelength of convection, i.e.,~with two counter-rotating rolls.  
Beyond the highest value of the wavenumber $a$ at a given value of the
Rayleigh number $R$, no finite-amplitude solution could be obtained. It
appears that at a  given value of $R$ 
the convection pattern that is realized from random initial conditions in
the case of a large $\Gamma$, say $\Gamma = 10$, is close to that
which maximizes $\overline{\Theta}(z=0.5)$ where the bar indicates the
$x$-average. For this reason, the quantity $H\equiv \beta/2 - T(0.5) =
-\overline{\Theta}(0.5)+(\tanh \gamma z)/\gamma$ has been plotted in
Fig.~1(b) for the same solutions as indicated in Fig.~1(a).  At the
preferred value of the wavenumber $a$, the quantity $H$ assumes a
minimum which is approximately indicated by the closest integer value
of $a$.   

In Fig.~2(a) the quantity $H$ has been plotted as a function of $R$
for different values of the Prandtl number $P$. 
Computations at much
higher values of $R$ appear to indicate that $H$ tends to zero for $R
\rightarrow \infty$. But these values of $R$ have not been included in
Fig.~2 because of insufficient numerical accuracy.
Corresponding values of $a$ are displayed in Fig.~2(b) which
indicates that the preferred $a$ does not vary much with the Prandtl
number $P$.  

\begin{figure}
\vspace*{1mm}
\begin{center}
\epsfig{file=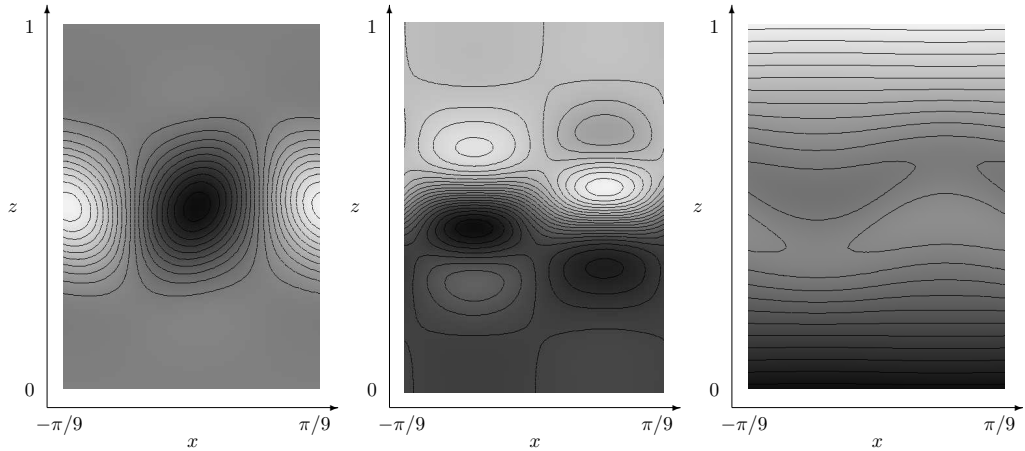,width=\textwidth,clip=}
\end{center}
\caption{Patterns of convection in the case $P=1$, $R=850000$,
  $\gamma=10$, $\beta=0.5$, corresponding to a single wavelength of the dominant
  wavenumber $a=9$. The first, second and  third plots show
  $\phi$, $\Theta$ and   $T=T_s+\Theta$, respectively. Whiter  shades
  correspond to negative  values and darker shades to positive values.
  }
\end{figure}

\begin{figure}
\vspace*{7mm}
\begin{center}
\epsfig{file=fig04.eps,width=\textwidth,clip=}
\end{center}
\caption{Time series of the kinetic energy in the case $P=0.1$,
  $R=850000$, $\gamma=10$, and $\beta=0.5$.
}
\vspace{7mm}
%
\begin{center}
\epsfig{file=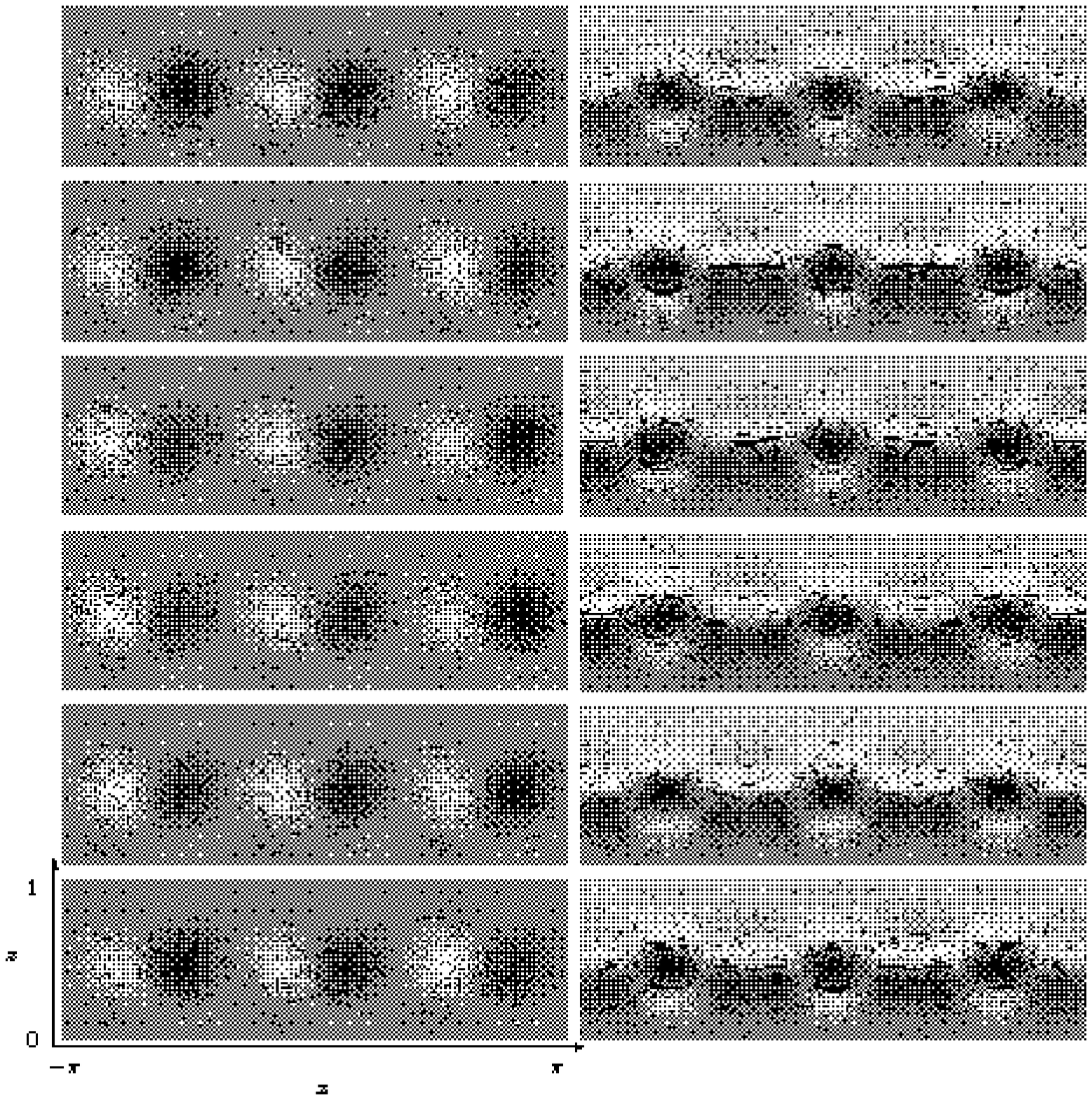,width=\textwidth,clip=}
\end{center}
\caption{A period of oscillations in the case shown in Fig.~4. The left
 and the right columns shows contour plots of $\phi$ and $\Theta$,
 respectively. The time interval between two plots is $0.32$ such that
 approximately one period is covered. 
}
\end{figure}

Typical convection patterns corresponding to the dominant wavenumber
of convection are shown in Fig.~3. These patterns are 
stationary due to the restriction of the horizontal size of the domain
to $\Gamma = 2\pi/a$, which forces the system to select a single
wavenumber. 
As the aspect ratio of the domain is increased, temporal and
spatial modulation of the patterns occur especially for large values
of the Rayleigh number $R$ and for smaller values of the Prandtl
number $P$. As an example, Figs.~4 and 5 illustrate a time-dependent 
modulation with the wavelength of $\Gamma$, which in this case is
set to $\Gamma=6\pi/a$. 

A most surprising phenomenon is exhibited in Fig.~6 where the mean
temperature, $T_s(z) + \overline{\Theta}$,  and the mean temperature
gradient have been plotted in 6(a) and 6(b), respectively. Unexpectedly,
the region of decreasing mean temperature with height characterizing
the onset of convection gives way to a purely increasing mean
temperature with height as the Rayleigh number exceeds about twice its
critical value. This effect is even more visible if the 
quantity $G\equiv\tanh(\gamma z)/\gamma  -\overline{\Theta}$ is
plotted as a function of the vertical coordinate $z$ as shown in
Fig.~7. Although a stably stratified layer is thus achieved in  
the mean, convection flows continue to be vigorous 
since they are driven by self-created horizontal temperature
differences. The results do not depend much on the value of the
Prandtl number.   

\begin{figure}
\vspace*{7mm}
\psfrag{T}{\hspace*{-4mm}$T_s+\overline{\Theta}$}
\psfrag{z}{$z$}
\psfrag{dT}{\hspace*{-4mm}$\partial_z  (T_s+\overline{\Theta})$}
\begin{center}
\epsfig{file=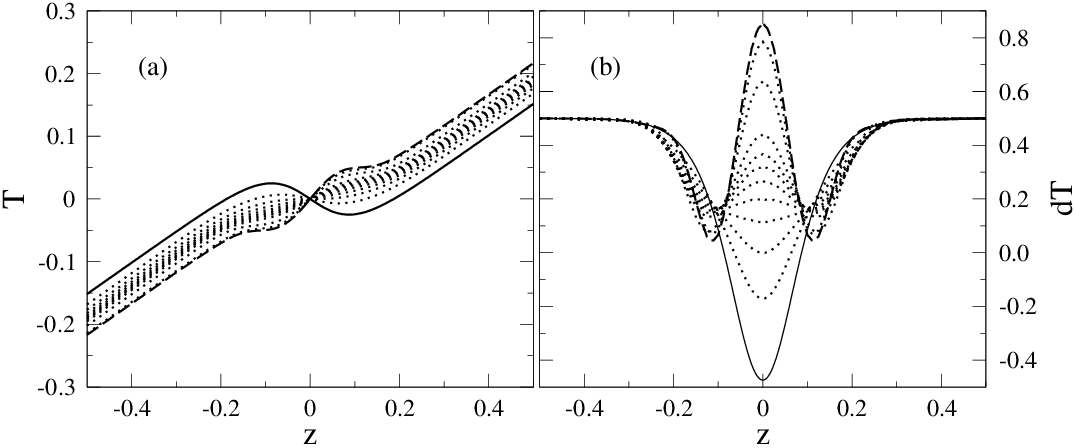,width=\textwidth,clip=}
\end{center}
\caption{(a) $T=T_s+\overline{\Theta}$ and (b) $\partial_z
  T=\partial_z
  (T_s+\overline{\Theta})$ as a function of the vertical
  coordinate $z$ for $P=10$, $\gamma=10$, $\beta=0.5$ and
  $R=150000+i\times10^5$, $i=0..7$, $R=$1050000, 2000000, 4000000, 6000000. The
  solid line indicates $R=150000$, 
  the broken line indicates $R=6000000$ and the dotted lines indicate
  the intermediate values of $R$.
}
\psfrag{T8}{$G$}
\psfrag{z}{$z$}
\vspace*{7mm}
\begin{center}
\hspace*{-1.5cm}
\epsfig{file=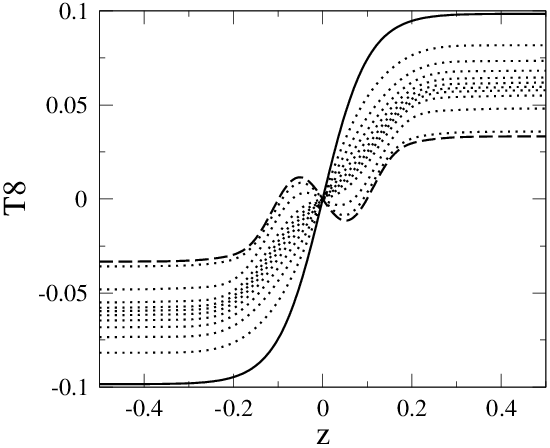,width=8cm,clip=}
\end{center}
\caption{The quantity $G\equiv\tanh(\gamma z)/\gamma
  -\overline{\Theta}$ as a function of the vertical
  coordinate $z$ with the same parameter values and linetypes as in
  Fig.~6.
}
\end{figure}


\section{Discussion and outlook}

Our study has been restricted in several aspects. We have focused on
an idealized case with antisymmetric heating (cooling) and have not
varied the parameters $\gamma$ and $\beta$. No comparison with
naturally occurring systems has been attempted. Most importantly, 
we have considered in this report only a two-dimensional
formulation. Further within this setting, we have performed the main
part of our simulations so as to confine the flow to one of the
possible roll wavenumbers. This has been done in order to
investigate the behavior of the system in its simplest
manifestation. We have found the unexpected effect that
convection is driven by lateral variations in temperature even at
moderate values of the Rayleigh number. We expect that this will
result in patterns and dynamics quite different from those familiar
from the case of the Rayleigh-B\'enard problem in horizontal layers
without stably stratified regions.
In particular, we expect to find complex time-dependent behavior, a
possibility clearly indicated by results such as the case presented
in Figs.~4 and 5, as well as by our preliminary three-dimensional
simulations, an example of which is shown in Fig.~8. Thus, a main
goal in the 3D-case  will be to study time dependences and the
possibility of interaction with internal waves in the stably
stratified regions of the layer. 

\begin{figure}
\vspace*{7mm}
\begin{center}
\epsfig{file=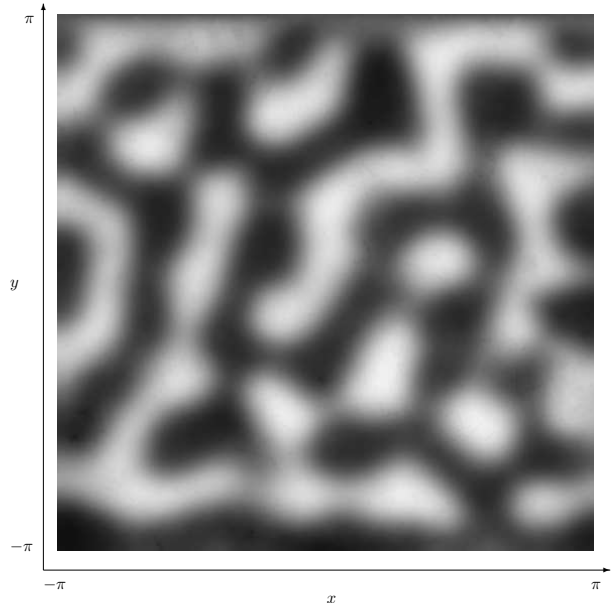,width=8cm,clip=}
\end{center}
\caption{Three-dimensional patterns of convection in the case $P\to\infty$,
  $R=10^6$,  $\beta=0.5$, $\gamma=10$ visualized by a grey-scale map
  of $\theta$ in the plane $z=0$. 
}
\end{figure}

\vspace*{0.2in}
\textbf{Acknowledgments} We gratefully acknowledge the support of CTR
and NASA which made possible our visits to Stanford. The COMSOL
software has been licensed to the School of Mathematics and Statistics
of the University of Glasgow, UK.

%


\end{document}